\documentclass[aps,prl,reprint,superscriptaddress]{revtex4-1}
\usepackage{amsmath}
\usepackage{amssymb}
\usepackage{amsthm}
\usepackage{graphicx,subfigure}
\usepackage[bookmarks=false]{hyperref}
\usepackage{bm,bbm}
\hypersetup{colorlinks=true,citecolor=blue,linkcolor=red,%
urlcolor=blue,pdfstartview=FitH,bookmarksopen=true}

\usepackage[T1]{fontenc}
\usepackage{txfonts}

\newtheorem*{theorem*}{Theorem}

\DeclareMathOperator{\Tr}{Tr}

\newcommand{\bra}[1]{\langle #1\rvert}
\newcommand{\ket}[1]{\lvert #1\rangle}

\newcommand{\abs}[1]{\lvert #1\rvert}
\newcommand{\norm}[1]{\lVert #1\rVert}
\newcommand{\vect}[1]{\bm{#1}}

\newcommand{\mi}{\mathrm{i}}

\newcommand{\I}{\mathbbm{1}}
\newcommand{\FA}{\text{\quad for all~~}}
\newcommand{\Exp}[1]{\mathrm{e}^{#1}}

\newcommand{\maxover}[1][]{\underset{#1}{\text{maximize}}}

\newcommand{\subto}{\text{subject to}}

\begin{document}

\title{Optimal verification of general bipartite pure states}

\author{Xiao-Dong Yu}
\email{Xiao-Dong.Yu@uni-siegen.de}
\affiliation{Naturwissenschaftlich-Technische Fakult\"at, Universit\"at Siegen,
Walter-Flex-Str. 3, D-57068 Siegen, Germany}

\author{Jiangwei Shang}
\email{jiangwei.shang@bit.edu.cn}
\affiliation{Beijing Key Laboratory of Nanophotonics and Ultrafine
Optoelectronic Systems, School of Physics, Beijing Institute of Technology,
Beijing 100081, China}
\affiliation{State Key Laboratory of Surface Physics and Department of Physics,
Fudan University, Shanghai 200433, China}

\author{Otfried G\"uhne}
\email{otfried.guehne@uni-siegen.de}
\affiliation{Naturwissenschaftlich-Technische Fakult\"at, Universit\"at Siegen,
Walter-Flex-Str. 3, D-57068 Siegen, Germany}

\date{\today}

\begin{abstract}
  The efficient and reliable verification of quantum states plays a crucial
  role in various quantum information processing tasks. We
  consider the task of verifying entangled states using one-way and two-way
  classical communication and completely characterize the optimal strategies
  via convex optimization.  We solve these optimization problems using both
  analytical and numerical methods, and the optimal strategies can be
  constructed for any bipartite pure state. Compared with the
  nonadaptive approach, our adaptive strategies significantly improve the
  efficiency of quantum state verification.  Moreover, these strategies are
  experimentally feasible, as only few local projective measurements are
  required.
\end{abstract}

\maketitle

\textit{Introduction.---}%
A basic yet important step in most quantum information processing tasks
is to efficiently and reliably characterize a quantum state. The standard
approach is to perform quantum state tomography by fully reconstructing the
density matrix \cite{Paris.Rehacek2004}. However, tomography is known to
be both time consuming and computationally hard due to the exponentially
increasing number of parameters to be reconstructed \cite{Haeffner.etal2005,
Shang.etal2017}; moreover, the underlying approximations may be conceptually
problematic \cite{Schwemmer.etal2015}. In fact, full tomographic information is
often not required, and a lot of effort has been devoted to characterizing
quantum states with non-tomographic methods
\cite{Toth.Guehne2005c,Gross.etal2010,Flammia.Liu2011,daSilva.etal2011}.
Recently, an alternative statistical approach, namely quantum state
verification, has triggered much research interest due to its powerful
efficacy \cite{Pallister.etal2018,Dimic.Dakic2018,Morimae.etal2017,
Takeuchi.Morimae2018,Zhu.Hayashi2018}.

Quantum state verification is a procedure for gaining confidence that
the output of some quantum device is a particular state by employing
local measurements \cite{Pallister.etal2018}. Consider a device that is
supposed to produce the target state $\ket{\psi}$, but may in
practice produce $\sigma_1,\sigma_2,\dots,\sigma_N$ in $N$ runs. In
the ideal scenario, the verifier has the promise that either
$\sigma_k=\ket{\psi}\bra{\psi}$ for all $k$ or
that $\sigma_k$ have a finite distance to $\ket{\psi}$, i.e.,
$\bra{\psi}\sigma_k\ket{\psi}\le 1-\varepsilon$ for all $k$. Given
access to some set of allowed measurements, the verifier must certify
that the source prepares $\ket{\psi}.$ One cannot exclude that
he certifies the source to be correct although it is not, but this
failure probability $\delta$ should be as small as possible.

In general, for each state $\sigma_k$ the verifier may apply a different
measurement with some predefined probability. So a state verification
strategy can be expressed as $\Omega=\sum_{i=1}^n p_i\Omega_i$,
where $(p_1,p_2,\dots,p_n)$ is a probability distribution, and
$\{\Omega_i,\I-\Omega_i\}$ are allowed measurements with outcomes labeled
by ``pass'' and ``fail'' respectively. For each output state $\sigma_k$, the
verifier randomly chooses a measurement $\{\Omega_i,\I-\Omega_i\}$ with
probability $p_i$, then performs the test.  In a pass instance, the verifier
continues to state $\sigma_{k+1}$, otherwise the verification ends and the
verifier concludes that the state was not $\ket{\psi}$. To guarantee
that the perfect state $\ket{\psi}$ is never rejected we assume
$\Omega_i$ satisfies $\bra{\psi}\Omega_i\ket{\psi}=1$;
it has been observed in Ref.~\cite{Pallister.etal2018} that
such strategies are better than others. The worst-case failure
probability of each run is given by
$\max_{\bra{\psi}\sigma\ket{\psi}\le 1-\varepsilon}\Tr(\Omega\sigma)
=1-\varepsilon v(\Omega)$,
where $v(\Omega)$ represents the spectral gap between the largest and the
second largest eigenvalues of $\Omega$
\cite{Pallister.etal2018}.

In the case that all $N$ states pass the test, we achieve the confidence
$1-\delta$ with
\begin{equation}
  \delta\le[1-\varepsilon v(\Omega)]^N.
  \label{eq:failureProbability}
\end{equation}
In reality, however, quantum devices are never perfect, so the verifier cannot
be promised that either $\sigma_k=\ket{\psi}\bra{\psi}$ or
$\bra{\psi}\sigma_k\ket{\psi}\le 1-\varepsilon$ for all $k$. Instead,
a more practical task is to certify with high confidence that the fidelity of
the output state is larger than a threshold value $1-\varepsilon$.
In this case, the verifier measures the frequency $f$ of the pass instances. If
$f>1-\varepsilon v(\Omega)$, the confidence $1-\delta$ can be derived
from the Chernoff bound \cite{Chernoff1952,Hoeffding1963}
\begin{equation}
 \delta\le\mathrm{e}^{-D[f \|(1-\varepsilon v(\Omega))]N},
\label{eq:entanglementVerification}
\end{equation}
where $D(x\|y)=x\log(\frac{x}{y})+(1-x)\log(\frac{1-x}{1-y})$ is the
Kullback-Leibler divergence.

The advantage of the state verification approach is that the failure
probability $\delta$ decreases exponentially with $N$, hence the target
state $\ket{\psi}$ can be potentially verified using only few copies of the
state. As seen from Eqs.~\eqref{eq:failureProbability} and
\eqref{eq:entanglementVerification}, the performance of a verification
strategy depends solely on $v(\Omega)$. Therefore, to achieve an optimal
strategy, we need to maximize $v(\Omega)$ over all accessible measurements.
Although lots of effort has been devoted to this research line, few optimal
strategies have been found. To the best of our knowledge, the only optimal
strategy reported by now is the verification of two-qubit pure states with
local projective measurements \cite{Pallister.etal2018}.

In this work, we introduce adaptive measurements, i.e., measurements
assisted by local operations and classical communication (LOCC)
\cite{Bennett.etal1996, Watrous2018} to the task of quantum state verification.
We show that the efficiency of the verification can be significantly improved
by considering adaptive measurements. For any $d_1\times d_2$ bipartite pure
state, we explicitly construct the optimal one-way as well as near-optimal
two-way adaptive verification strategies.  Best of all, these strategies are
experimentally friendly as only few local projective measurements are needed
for their implementation in the laboratory.

\textit{Optimal state verification as convex optimization.---}%
In the following, we derive two convex optimization problems that completely
characterize the optimal adaptive state verification strategies assisted by
one-way and one-round two-way classical communication respectively. In general,
to get an optimal verification strategy, we need to consider the optimization
problem
\begin{equation}
  \begin{aligned}
    &\maxover[p_i,\Omega_i] \quad &&v(\Omega)\\
    &\subto                       &&\Omega=\sum_{i=1}^np_i\Omega_i,\\
    &                             &&\sum_{i=1}^np_i=1,~~p_i\ge 0,\FA i,\\
    &                             &&\bra{\psi}\Omega_i\ket{\psi}=1,
				  ~~\Omega_i\in\mathcal{M},\FA i,
  \end{aligned}
  \label{eq:optimizationGeneral}
\end{equation}
where $\ket{\psi}$ is the target state we want to verify, and
$\mathcal{M}$ denotes the set of all allowed measurements.  Be reminded that
$v(\Omega)$ represents the spectral gap
between the largest and the second largest eigenvalues of $\Omega$. As
$\Omega_i\le\I$, the last constraint 
leads to $\Omega_i\ket{\psi}=\ket{\psi}$ and
$P^\perp\Omega_i P^\perp=\Omega_i-\ket{\psi}\bra{\psi}$, where
$P^\perp=\I-\ket{\psi}\bra{\psi}$. Hence, $v(\Omega)$ admits an
alternative expression
\begin{equation}
  v(\Omega)=1-\norm{P^\perp\Omega P^\perp},
  \label{eq:vOmega}
\end{equation}
where $\norm{\,\cdot\,}$ denotes the largest eigenvalue.

Generally speaking, the optimization in Eq.~\eqref{eq:optimizationGeneral} is
difficult to solve, if not impossible at all, because the set of all possible
measurements cannot be easily characterized.  Here, we give
a complete characterization of $\Omega$ for both one-way and one-round two-way
adaptive measurements, then reduce the corresponding problems to convex optimization.
These optimization problems can be further simplified and solved.
For succinctness, hereafter we restrict the two-way adaptive measurements to
one-round communication only. In addition,
the accessible measurements allowed in our verification strategies are not
restricted to projective measurements (PMs), i.e., positive operator-valued
measures (POVMs) are possible, although in the end we show that the optimal
strategies can be achieved with PMs in most cases.

Without loss of generality, a bipartite pure state can be written as
$\ket{\psi}=\sum_{i=1}^d\lambda_i\ket{ii}$, where the Schmidt coefficients
satisfy $\lambda_1\ge\lambda_2\ge\dots\lambda_d>0$ and
$\sum_{i=1}^d\lambda_i^2=1$ \cite{Peres2002}.

We start with the analysis of one-way communication.
In this case, Alice first performs a measurement, and sends the
measurement outcome to Bob.  Bob then chooses his measurement
in accordance with Alice's measurement outcome. Hence, the one-way adaptive
strategy $\Omega^\rightarrow$ takes the form
\begin{equation}
  \Omega^\rightarrow=\sum_{i=1}^np_i\Omega_i^\rightarrow,\quad
  \Omega_i^\rightarrow=\sum_aM_{a|i}\otimes N_{a|i},
  \label{eq:adaptiveOneWayMeasurement}
\end{equation}
where $\{M_{a|i}\}_a$ are measurements on Alice's system, and each
$\{N_{a|i},\I-N_{a|i}\}$ is a ``pass'' or ``fail'' measurement on Bob's system
depending on Alice's measurement outcome. Here, we can assume that the
$M_{a|i}$ are rank-one, otherwise some further decomposition can make this
assumption satisfied. If the joint system is in state $\ket{\psi}$, Bob's
subsystem would collapse to some pure state
${P_{a|i}=\Tr_A(M_{a|i}\otimes\I\ket{\psi}\bra{\psi})/
\Tr(M_{a|i}\otimes\I\ket{\psi}\bra{\psi})}$
after Alice's measurement $\{M_{a|i}\}_a$.
Then the best strategy for Bob is to perform the measurement
$\{P_{a|i},\I-P_{a|i}\}$ to verify whether his subsystem is in
state $P_{a|i}$.  Mathematically, to ensure that
$\bra{\psi}\Omega_i^\rightarrow\ket{\psi}=1$, $N_{a|i}$ must satisfy that
$N_{a|i}\ge P_{a|i}$. If all $N_{a|i}$ satisfy $N_{a|i}=P_{a|i}$, we call
the one-way adaptive strategy $\Omega^\rightarrow$ semi-optimal.
Hence, to maximize $v(\Omega^\rightarrow)$, i.e.,
to minimize $\norm{\sum_ip_iP^\perp\Omega_i^\rightarrow P^\perp}$,
we can restrict $\Omega^\rightarrow$ to be semi-optimal strategies.

From the definition, we get the following necessary conditions for
$\Omega^\rightarrow$ being semi-optimal
\begin{equation}
  \Omega^\rightarrow\in\mathcal{S},\quad
  \Tr_B(\Omega^\rightarrow)=\I,\quad
  \bra{\psi}\Omega^\rightarrow\ket{\psi}=1,
  \label{eq:constraints}
\end{equation}
where $\mathcal{S}$ is the set of separable operators, i.e., unnormalized
separable states \cite{Watrous2018}.
Next, we show that these constraints
are also sufficient. $\Omega^\rightarrow$ is separable implies
that there exists a decomposition $\Omega^\rightarrow=\sum_aM_a\otimes N_a$,
such that $M_a$ are positive semidefinite and $N_a$ are rank-one projectors.
Then, $\Tr_B(\Omega^\rightarrow)=\I$ implies $\sum_aM_a=\I$, i.e., $\{M_a\}_a$
is a measurement on Alice's system. This concludes our proof by taking into
account the last constraint. Thus, the optimization in
Eq.~\eqref{eq:optimizationGeneral} can be written as
\begin{equation}
  \begin{aligned}
    &\maxover[\Omega^\rightarrow] \quad &&v(\Omega^\rightarrow)\\
    &\subto                       &&\Omega^\rightarrow\in\mathcal{S},\\
    &                             &&\Tr_B(\Omega^\rightarrow)=\I,\\
    &                             &&\bra{\psi}\Omega^\rightarrow\ket{\psi}=1,
  \end{aligned}
  \label{eq:optimizationOneWayGeneral}
\end{equation}
for one-way adaptive verification strategies.

We move on to discuss the one-round two-way communication scenario.
In this case, Alice and Bob use shared randomness to decide who performs
the measurement first. After the measurement, he/she sends the measurement
outcome to the other party. Then the receiver chooses her/his measurement
according to the received measurement outcome. Thanks to
the permutation symmetry of $\ket{\psi}$, the optimization in this setting can
be easily simplified. Let $S$ be the SWAP operator, i.e.,
$S\ket{i}\ket{j}=\ket{j}\ket{i}$ for all $i,j=1,2,\dots,d$,
then we have $S\ket{\psi}=\ket{\psi}$. This indicates
that, for two-way adaptive measurements, if $\Omega$ satisfies the constraints
in Eq.~\eqref{eq:optimizationGeneral}, so does $\frac1{2}(\Omega+S\Omega S^\dagger)$.
Furthermore, Eq.~\eqref{eq:vOmega} implies
\begin{equation}
  v\left[\tfrac{1}{2}(\Omega+S\Omega S^\dagger)\right]\ge
  \tfrac{1}{2}\left[v(\Omega)+v(S\Omega S^\dagger)\right]=v(\Omega).
  \label{eq:largerMinSwap}
\end{equation}
Hence, we can focus on the two-way adaptive strategies $\Omega^\leftrightarrow$
that are invariant under the SWAP operation, i.e.,
${\Omega^\leftrightarrow=\frac{1}{2}(\Omega^\rightarrow+\Omega^\leftarrow)}$,
where $\Omega^\rightarrow$ is a one-way adaptive strategy and
$\Omega^\leftarrow=S\Omega^\rightarrow S^\dagger$. Similarly, to optimize
$v(\Omega^\leftrightarrow)$, we can also restrict $\Omega^\rightarrow$ to be
semi-optimal. Thus, the optimization
in Eq.~\eqref{eq:optimizationGeneral} can be written as
\begin{equation}
  \begin{aligned}
    &\maxover[\Omega^\rightarrow] \quad &&v\left[\tfrac{1}{2}
			(\Omega^\rightarrow+\Omega^\leftarrow)\right]\\
    &\subto                       &&\Omega^\rightarrow\in\mathcal{S},\\
    &                             &&\Tr_B(\Omega^\rightarrow)=\I,\\
    &                             &&\bra{\psi}\Omega^\rightarrow\ket{\psi}=1,
  \end{aligned}
  \label{eq:optimizationTwoWayGeneral}
\end{equation}
for two-way adaptive verification strategies.

\textit{Optimal verification of two-qubit states.---}%
Without loss of generality, we write the two-qubit entangled pure state as
$\ket{\psi}=\cos\theta\ket{00}+\sin\theta\ket{11}$ with $0<\theta\le\pi/4$.
Then the subspace $P^\perp$ is
spanned by $\{\ket{\psi_i}\}_{i=1}^3:=\{\ket{01},\ket{10},
\sin\theta\ket{00}-\cos\theta\ket{11}\}$.

First, we need a group $G$ to simplify the optimizations.
The group $G$ is defined to be generated by the
unitary operator $g=\Phi\otimes\Phi^\dagger$, where $\Phi$ is the phase gate,
i.e., $\Phi\ket{0}=\ket{0}$ and $\Phi\ket{1}=\mi\ket{1}$.
Then we can show
\begin{equation}
  \tilde\Omega:=\frac{1}{4}\sum_{k=0}^3g^k\Omega g^{-k}
  =\sum_{i=1}^3w_i\ket{\psi_i}\bra{\psi_i}
  +\ket{\psi}\bra{\psi};
  \label{eq:diagonalOmega}
\end{equation}
see Appendix A for the proof.
As $g\ket{\psi}=\ket{\psi}$, $\tilde\Omega$ also satisfies the
constraints in Eqs.~\eqref{eq:optimizationOneWayGeneral} and
\eqref{eq:optimizationTwoWayGeneral} if $\Omega$ does. Furthermore,
Eq.~\eqref{eq:vOmega} implies
\begin{equation}
  v(\tilde\Omega)\ge\frac{1}{4}\sum_{k=0}^3
  v\left(g^k\Omega g^{-k}\right)=v(\Omega).
  \label{eq:largerMin}
\end{equation}
Thus, we can restrict to the diagonal $\Omega^\rightarrow$ as in
Eq.~\eqref{eq:diagonalOmega} for the optimizations in
Eqs.~\eqref{eq:optimizationOneWayGeneral} and
\eqref{eq:optimizationTwoWayGeneral}.

Then, we consider the case of one-way adaptive verification. For two-qubit
quantum states, the positive partial transpose (PPT) criterion is necessary and
sufficient to characterize their separability
\cite{Peres1996,Horodecki.etal1996}. Thus, by combining
Eq.~\eqref{eq:diagonalOmega} with the PPT criterion, the optimization
in Eq.~\eqref{eq:optimizationOneWayGeneral} can be written as
\begin{equation}
  \begin{aligned}
    &\maxover[w_i]\quad &&\min\{1-w_i\}\\
    &\subto  &&w_i\ge 0,~~i=1,2,3,\\
    &        &&w_1=\sin^2\theta(1-w_3),\\
    &        &&w_2=\cos^2\theta(1-w_3),
  \end{aligned}
  \label{eq:optimizationSimplified}
\end{equation}
where the constraints arise only from $\Omega^\rightarrow\ge0$ and
$\Tr_B(\Omega^\rightarrow)=\I$, since
the PPT criterion gives the redundant condition
$w_1w_2\ge\sin^2\theta\cos^2\theta(1-w_3)^2$. As $0<\theta\le\pi/4$, we have
$w_2\ge w_1$. Thus, the solution of Eq.~\eqref{eq:optimizationSimplified} is
attained when $w_2=w_3$, and
\begin{equation}
  \max_{\Omega^\rightarrow}v(\Omega^\rightarrow)
  =\frac{1}{1+\cos^2\theta}.
  \label{eq:solutionOneWayCC}
\end{equation}

In general, the measurements associated with the optimal solution are POVMs.
However, one can directly calculate that the bound in
Eq.~\eqref{eq:solutionOneWayCC} can be achieved already with PMs
\begin{equation}
  \Omega^{\rightarrow}=\frac{\cos^2\theta}{1+\cos^2\theta}P_{ZZ}^+
  +\frac{1}{2(1+\cos^2\theta)}X_\psi^{\rightarrow}
  +\frac{1}{2(1+\cos^2\theta)}Y_\psi^{\to},
  \label{eq:OneWayPVM}
\end{equation}
where
\begin{equation}
  \begin{aligned}
    P_{ZZ}^+&=\ket{0}\bra{0}\otimes\ket{0}\bra{0}
    +\ket{1}\bra{1}\otimes\ket{1}\bra{1},\\
    X_\psi^{\rightarrow}&=\ket{\varphi_0}\bra{\varphi_0}
    +\ket{\varphi_2}\bra{\varphi_2},\\
    Y_\psi^{\rightarrow}&=\ket{\varphi_1}\bra{\varphi_1}
    +\ket{\varphi_3}\bra{\varphi_3},
  \end{aligned}
  \label{eq:defXYZ}
\end{equation}
with
$\ket{\varphi_0}=\frac{1}{\sqrt{2}}(\ket{0}+\ket{1})
\otimes(\cos\theta\ket{0}+\sin\theta\ket{1})$ and
$\ket{\varphi_k}=g^k\ket{\varphi_0}$.

Next, we discuss the case of two-way adaptive verification. By combining
Eq.~\eqref{eq:diagonalOmega} and the PPT criterion, we can get a simplification
of the optimization in
Eq.~\eqref{eq:optimizationTwoWayGeneral} by simply replacing the objective
function in Eq.~\eqref{eq:optimizationSimplified} with
\begin{equation}
  \maxover[w_i]\quad \min\left\{1-\tfrac{1}{2}(w_1+w_2),1-w_3\right\},
  \label{eq:optimizationSimplifiedTwo}
\end{equation}
whose solution is given by
\begin{equation}
  \max_{\Omega^\leftrightarrow}v(\Omega^\leftrightarrow)
  =\frac{2}{3}.
  \label{eq:solutionTwoWayCC}
\end{equation}
Again, we explicitly write down the PMs
\begin{equation}
    \Omega^{\leftrightarrow}=\frac{1}{3}P_{ZZ}^++\frac{1}{6}X_\psi^{\rightarrow}
    +\frac{1}{6}X_\psi^{\leftarrow}+\frac{1}{6}Y_\psi^{\rightarrow}
    +\frac{1}{6}Y_\psi^{\leftarrow},
  \label{eq:adptativeMeasurement}
\end{equation}
where $P_{ZZ}^+$, $X_\psi^{\rightarrow}$, and $Y_\psi^{\rightarrow}$ are
defined as in Eq.~\eqref{eq:defXYZ}, and
$X_\psi^{\leftarrow}=SX_\psi^{\rightarrow}S^{\dagger}$ and
$Y_\psi^{\leftarrow}=SY_\psi^{\rightarrow}S^{\dagger}$.

\begin{figure}[t]
  \includegraphics[width=.85\columnwidth]{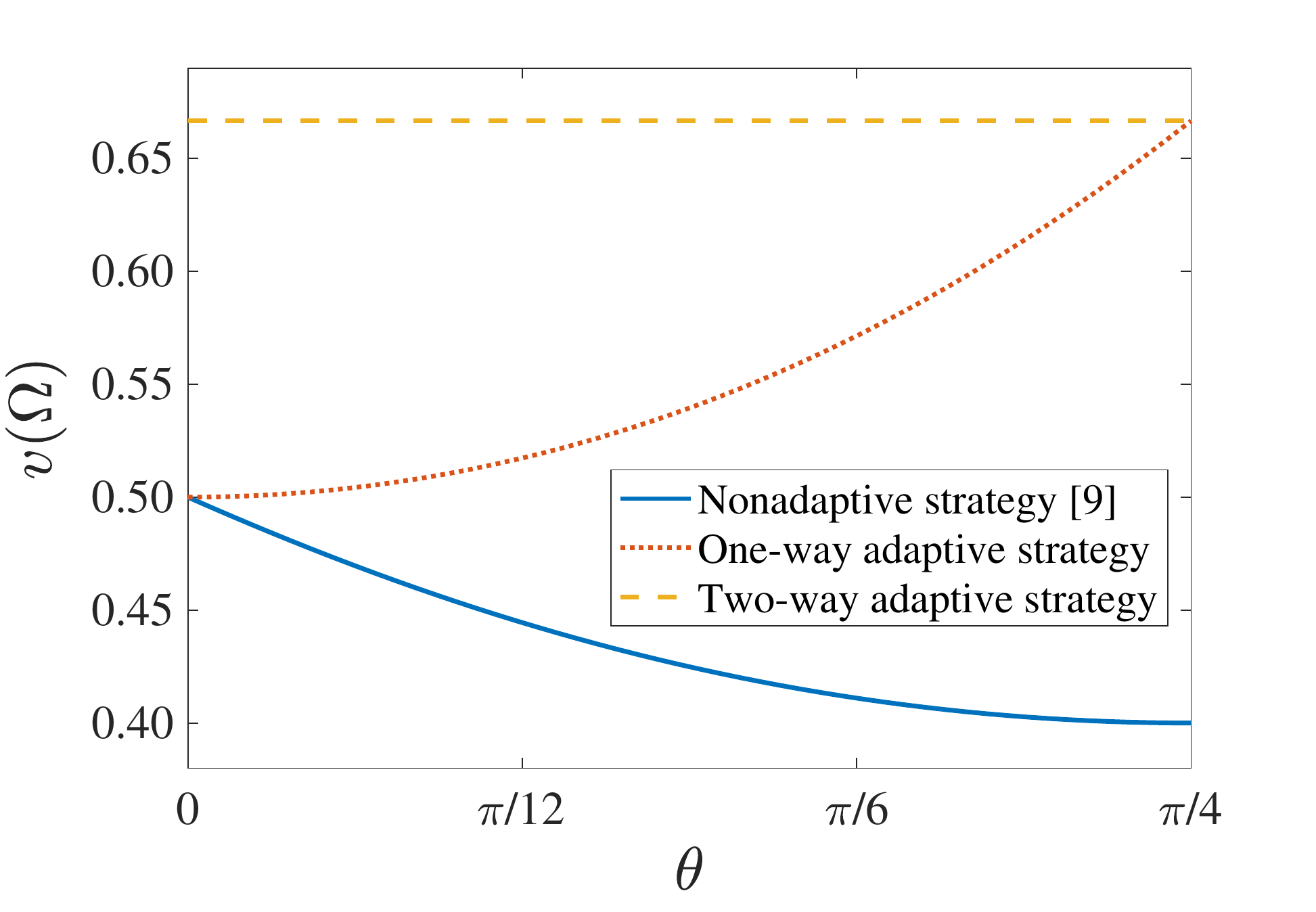}
  \caption{Optimal values of $v(\Omega)$ with different verification strategies
    for the two-qubit entangled pure state
    $\ket{\psi}=\cos{\theta}\ket{00}+\sin{\theta}\ket{11}$ with
    $0<\theta<{\pi}/{4}$.
    Note that when $\theta={\pi}/{4}$, i.e., $\ket{\psi}$ is the maximally
    entangled state,
    all three strategies give the same optimal value $v(\Omega)=2/3$.}
  \label{fig:comparison}
\end{figure}

Finally, we compare the adaptive strategies with the nonadaptive approach in
Ref.~\cite{Pallister.etal2018}. For two-qubit entangled states, we plot the
optimal values of $v(\Omega)$ for different strategies in
Fig.~\ref{fig:comparison}.  As can be seen, the two-way strategy works
much better than the one-way strategy, whereas both the adaptive strategies
significantly outperform the nonadaptive one. Concerning the resources used in
each strategy, we have the following remarks.
Although no classical communication is
involved in the measurement process of the nonadaptive strategy, it is still
a necessary resource for the data processing after the measurement.  On the
contrary, the one-way adaptive strategy relies on classical
communication for the measurements, but no classical
communication is needed for the
data processing as one party alone can determine whether the result is
a pass or fail instance. The case for the two-way adaptive strategy is similar,
but to obtain the final frequency of the pass instances, the two parties need
to cooperate.

\textit{Optimal verification of general bipartite states.---}%
We move on to discuss the optimal adaptive verification of general bipartite
states. Firstly, we need a larger group $G$ for the general bipartite (two-qudit)
pure state $\ket{\psi}=\sum_{i=1}^d\lambda_i\ket{ii}$.
The group $G$ is defined to be generated by the unitary
operators $\{g_k=\Phi_k\otimes\Phi_k^\dagger,~k=1,2,\dots,d\}$,
where $\Phi_k\ket{j}=\mi\ket{j}$ when $j=k$, and
$\Phi_k\ket{j}=\ket{j}$ otherwise.
Then we can show
\begin{equation}
  \tilde\Omega:=\frac{1}{\abs{G}}\sum_{g\in G}g\Omega g^\dagger
  =\sum_{j\ne i,i=1}^dw_{ij}\ket{ij}\bra{ij}
  +\sum_{i,j=1}^d\rho_{ij}\ket{ii}\bra{jj},
  \label{eq:symmetrizedQudit}
\end{equation}
where $\abs{G}$ is the order of $G$; see Appendix A for the proof.
Similar to the two-qubit case, if $\Omega$ satisfies the constraints in
Eqs.~\eqref{eq:optimizationOneWayGeneral} and
\eqref{eq:optimizationTwoWayGeneral}, so does $\tilde\Omega$, since
$g\ket{\psi}=\ket{\psi}$ for all $g\in G$. Furthermore, Eq.~\eqref{eq:vOmega}
implies
\begin{equation}
  v(\tilde\Omega)\ge\frac{1}{\abs{G}}\sum_{g\in G}
  v\left(g\Omega g^\dagger\right)=v(\Omega).
  \label{eq:largerMinQudit}
\end{equation}
Hence, we can restrict $\Omega^\rightarrow$ to
be of the form in Eq.~\eqref{eq:symmetrizedQudit} for the optimizations in
Eqs.~\eqref{eq:optimizationOneWayGeneral} and
\eqref{eq:optimizationTwoWayGeneral}. Additionally,
$\bra{\psi}\Omega\ket{\psi}=1$, i.e., $\Omega\ket{\psi}=\ket{\psi}$, means
$\rho\vect{\lambda}=\vect{\lambda}$,
where $\rho:=(\rho_{ij})_{i,j=1}^d$ is Hermitian, and
$\vect{\lambda}=(\lambda_1,\lambda_2,\dots,\lambda_d)^T$.

Secondly, we consider the case of one-way adaptive verification.  The main
difference between two-qudit and two-qubit states is that the PPT criterion is
only necessary but not sufficient to characterize the separability for
$d\ge 3$ \cite{Horodecki.etal1996}. Hence, by replacing
$\Omega^\rightarrow\in\mathcal{S}$ with $(\Omega^\rightarrow)^{T_B}\ge 0$,
Eqs.~\eqref{eq:optimizationOneWayGeneral} and \eqref{eq:symmetrizedQudit} only
give us a relaxation of the original optimization
\begin{equation}
  \begin{aligned}
    &\maxover[w_{ij},~\rho_{ij}] \quad &&\min\left\{1-w_{ij},
    1-\norm{\rho-\vect{\lambda}\vect{\lambda}^T}\right\}\\
    &\subto  &&0\le\rho\le\I,~~w_{ij}\ge 0,\FA i\ne j,\\
    &        &&w_{ij}w_{ji}\ge\abs{\rho_{ij}}^2,\FA i\ne j,\\
    &        &&\sum_{j\ne i}w_{ij}+\rho_{ii}=1,\FA i,\\
    &        &&\rho\vect{\lambda}=\vect{\lambda},
  \end{aligned}
  \label{eq:relaxationOptimizationOneWayCCQudit}
\end{equation}
where the constraints arise from $0\le\Omega^\rightarrow\le\I$,
the PPT criterion, $\Tr_B(\Omega^\rightarrow)=\I$, and
$\bra{\psi}\Omega^\rightarrow\ket{\psi}=1$ respectively.
Therefore, the solution of this relaxed problem
sets an upper bound
of the optimal $v(\Omega^\rightarrow)$. To show that the solution is a valid
strategy, we still need to prove that the optimal $\Omega^\rightarrow$ obtained
from Eq.~\eqref{eq:relaxationOptimizationOneWayCCQudit} is separable.
Here, instead of resorting to numerical methods, we can analytically solve the
optimization in Eq.~\eqref{eq:relaxationOptimizationOneWayCCQudit},
which gives
\begin{equation}
  \max_{\Omega^\rightarrow}v(\Omega^\rightarrow)\le\frac{1}{1+\lambda_1^2},
  \label{eq:simplifiedOptimalSolutionOneWay}
\end{equation}
for all $d\ge 2$. Moreover, the bound
in Eq.~\eqref{eq:simplifiedOptimalSolutionOneWay} can be achieved with PMs
\begin{equation}
  \Omega^\rightarrow=wP_{ZZ}+\frac{1-w}{\abs{G}}\sum_{g\in
  G}gX_\psi^\rightarrow g^\dagger,
  \label{eq:oneWayPVMQudit}
\end{equation}
where
\begin{equation}
  \begin{aligned}
    &P_{ZZ}=\sum_{k=1}^d\ket{k}\bra{k}\otimes\ket{k}\bra{k},
    &&X_\psi^\rightarrow=\sum_{k=1}^d\ket{f_k}\bra{f_k}
    \otimes\ket{\phi_k}\bra{\phi_k},\\
    &\ket{f_k}=\frac{1}{\sqrt{d}}\sum_{j=1}^d\gamma_d^{jk}\ket{j},
    &&\ket{\phi_k}=\sum_{j=1}^d\gamma_d^{-jk}\lambda_j\ket{j},
  \end{aligned}
  \label{eq:fourierBasis}
\end{equation}
with $\gamma_d=\Exp{\frac{2\pi\mi}{d}}$ and $w=\lambda_1^2/(1+\lambda_1^2)$;
see Appendix B for more details. In passing, we note two
special cases of Eq.~\eqref{eq:fourierBasis}. When $\ket{\psi}$ is separable,
i.e., $d=1$, Eq.~\eqref{eq:fourierBasis} gives the optimal nonadaptive strategy
with $v(\Omega)=1$. When $\ket{\psi}$ is maximally entangled,
$\{\ket{\phi_k}\}_{k=1}^d$ forms an orthogonal basis. Hence,
Eq.~\eqref{eq:fourierBasis} gives the optimal nonadaptive strategy
\cite{Pallister.etal2018,Zhu.Hayashi2019}.

In practice, the above strategy can be easily implemented.
Alice first randomly chooses one of the two measurements $\{\ket{k}\}_{k=1}^d$ and
$\{\ket{f_k}\}_{k=1}^d$ with probabilities $w$ and $1-w$ respectively.
The former measurement can be performed directly, while the latter one
requires some random phase shifts from $G$ in advance.  Then Alice sends all
the information to Bob via classical communication, upon receiving which Bob
can proceed to perform the corresponding test.

Lastly, we consider the case of two-way adaptive verification.
By the same token, the efficiency can be improved by averaging
$\Omega^\rightarrow$ and its swap $\Omega^\leftarrow$. Specifically, we can
get
\begin{equation}
  v(\Omega^\leftrightarrow)=v\left[\tfrac{1}{2}(\Omega^\rightarrow
  +\Omega^\leftarrow)\right]=\frac{1}{1+\lambda^2},
  \label{eq:nearOptimalSolutionTwoWay}
\end{equation}
when $\Omega^\rightarrow$ is of the form in Eq.~\eqref{eq:oneWayPVMQudit} with
$w=\lambda^2/(1+\lambda^2)$  and
$\lambda^2=\frac1{2}(\lambda_1^2+\lambda_2^2)$. However, unlike the two-qubit
case, this strategy is only near-optimal for general bipartite states.  To get
the optimal strategy, we can numerically solve the optimization in
Eq.~\eqref{eq:nearOptimalSolutionTwoWay}, then explicitly decompose
the obtained strategy with the method in Ref.~\cite{Shang.Guehne2018}.
Our testing results show that the optimal strategy is at most $4\%$ better in efficiency than
the near-optimal strategy for all $d\le 10$, whereas the measurement settings
can be more complicated; see Appendix C for more details.

Before concluding, two remarks are in order.
First, Eqs.~\eqref{eq:simplifiedOptimalSolutionOneWay} and
\eqref{eq:nearOptimalSolutionTwoWay} imply that $v(\Omega)\ge 1/2$ for all of
our adaptive strategies. This implies that
$N\lesssim 2\varepsilon^{-1}\log\delta^{-1}$ copies of states are enough for
verifying any bipartite states, which is
independent of the dimension $d$. This is of the same scale with the best
global strategies with entangled measurements, which need
$N\approx\varepsilon^{-1}\log\delta^{-1}$ copies \cite{Pallister.etal2018}. On
the contrary, the best nonadaptive strategies known so far need $N\gtrsim
d\varepsilon^{-1}\log\delta^{-1}$ to verify a generic two-qudit state for $d\ge
3$ \cite{Liu.etal2019}, which is worse than our adaptive strategies by an order
$O(d)$.  Second, it is possible to further improve the efficiency of the
adaptive strategies by involving many-round communication
\cite{Wang.Hayashi2019}. However, these strategies require coherence-preserving
measurements and can only improve the efficiency up to a constant factor $c$
with $c\le 2$ for all dimensions.

\textit{Conclusions.---}%
Quantum state verification is an efficient and reliable method
for gaining confidence about the quality of quantum devices,
which is a crucial step in almost all quantum information processing
tasks and foundational studies. In this work, we integrated
adaptive measurements to the problem of state verification and formulated
two convex optimization problems that completely characterize the optimal
adaptive strategies for one-way and one-round two-way classical communication.
We solve these optimization problems using both analytical and numerical
methods, and the optimal or near-optimal strategies are constructed explicitly
for any bipartite pure state. As a demonstration, we compared
the optimal adaptive strategies with the nonadaptive one, and find that the
verification efficiency can be significantly improved if classical
communication is allowed.  Finally, our adaptive verification strategies are
readily applicable in experiments as only few local projective measurements are
involved.  For future research, it is very interesting to consider the
multipartite case, which is more relevant for applications.
Moreover, it is meaningful to discuss how the present approach needs
to be modified, if the measurement devices are not perfectly characterized.
Statistical tools developed for quantum state discrimination
\cite{Barnett.Croke2009,Bae.Kwek2015} may be helpful for this purpose.

\begin{acknowledgments}
  We would like to thank Mariami Gachechiladze and Chau Nguyen for discussions.
  This work was supported by  the DFG and the ERC (Consolidator Grant
  683107/TempoQ). X.D.Y. acknowledges funding from a CSC-DAAD scholarship. J.S.
  acknowledges support by the Beijing Institute of Technology Research Fund
  Program for Young Scholars and the National Natural Science Foundation of
  China through Grant No. 11805010.

  \textit{Note added.---}During the preparation of the manuscript we became aware
  of related works by Wang and Hayashi \cite{Wang.Hayashi2019}, and Li
  \textit{et al.} \cite{Li.etal2019}.
\end{acknowledgments}

\appendix
%
\section{Appendix A: Proofs of Equations~\eqref{eq:diagonalOmega} and
\eqref{eq:symmetrizedQudit}}

It is easy to see that Eq.~\eqref{eq:diagonalOmega} is a special case of
Eq.~\eqref{eq:symmetrizedQudit} when $d=2$. Hence, we just need to prove Eq.~\eqref{eq:symmetrizedQudit}, which we restate below
\begin{equation}
  \tilde\Omega:=\frac{1}{\abs{G}}\sum_{g\in G}g\Omega g^\dagger
  =\sum_{j\ne i,i=1}^dw_{ij}\ket{ij}\bra{ij}
  +\sum_{i,j=1}^d\rho_{ij}\ket{ii}\bra{jj},
  \label{eq:symmetrizedQuditA}
\end{equation}
where $\abs{G}$ is the number of elements in group $G$. Recall that $G$ is
defined to be generated by
\begin{equation}
  g_k=\Phi_k\otimes\Phi_k^\dagger, \FA k=1,2,\dots,d,
  \label{eq:generatorQuditA}
\end{equation}
where $\Phi_k$ satisfies
\begin{equation}
 \begin{aligned}
   \Phi_k\ket{j}=
   \begin{cases}
     \mi\ket{j}\quad& j=k,\\
     \ket{j}\quad& j\ne k.
   \end{cases}
 \end{aligned}
 \label{eq:defPhaseGateA}
\end{equation}
We also note that
\begin{equation}
  \begin{aligned}
    w_{ij}&=\bra{ij}\tilde\Omega\ket{ij}=\bra{ij}\Omega\ket{ij},\\
    \rho_{ij}&=\bra{ii}\tilde\Omega\ket{jj}=\bra{ii}\Omega\ket{jj},
  \end{aligned}
  \label{eq:invariantBasis}
\end{equation}
as $\ket{ij}\bra{ij}$ and $\ket{ii}\bra{jj}$ are invariant under the group
action $G$.
To prove Eq.~\eqref{eq:symmetrizedQuditA}, we just need to show
\begin{equation}
  \bra{kl}\tilde{\Omega}\ket{ij}=0
  \label{eq:zeroEntries}
\end{equation}
unless
\begin{equation}
  \begin{aligned}
    &k=l,~i=j, \quad \text{or}
    &k=i,~l=j.
  \end{aligned}
  \label{eq:complement}
\end{equation}
Note that for all $g\in G$, we have
\begin{equation}
  \tilde{\Omega}=\frac{1}{2}(\tilde{\Omega}+g^\dagger\tilde{\Omega}g),
  \label{eq:average}
\end{equation}
because $\tilde{\Omega}$ is invariant under the group action, i.e.,
$g\tilde{\Omega}g^\dagger=\tilde{\Omega}$. To prove Eq.~\eqref{eq:zeroEntries},
we classify the quadruple $(k,l,i,j)$ into two different cases.

\textbf{Case 1:} Certain index in $(k,l,i,j)$ appears only once.  Without loss
of generality, we assume it is $k$. In this case, we choose $g=g_k^2$, then
\begin{equation}
  \begin{aligned}
    g\ket{kl}&=g_k^2\ket{kl}=-\ket{kl},\\
    g\ket{ij}&=g_k^2\ket{ij}=\ket{ij}.
  \end{aligned}
  \label{eq:group1}
\end{equation}
Combining with Eq.~\eqref{eq:average}, we obtain
$\bra{kl}\tilde{\Omega}\ket{ij}=0$.

\textbf{Case 2:} All indexes in $(k,l,i,j)$ appear more than once. Then the only
possibility excluded from Eq.~\eqref{eq:complement} is
\begin{equation}
  k=j\ne l=i.
  \label{eq:exception}
\end{equation}
In this case, we choose $g=g_k$, then
\begin{equation}
  \begin{aligned}
    g\ket{kl}&=g_k\ket{kl}=\mi\ket{kl},\\
    g\ket{lk}&=g_k^2\ket{lk}=-\mi\ket{lk}.
  \end{aligned}
  \label{eq:group2}
\end{equation}
Again, together with Eq.~\eqref{eq:average}, we get
$\bra{kl}\tilde{\Omega}\ket{ij}=0$.
This concludes the proof.

\section{Appendix B: Optimal one-way strategy}

In this appendix, we solve the optimization in
Eq.~\eqref{eq:relaxationOptimizationOneWayCCQudit}.
To illustrate the main idea behind our method, we first consider a special case
in which $\rho$ is of the form
\begin{equation}
  \rho=w\I+(1-w)\vect{\lambda}\vect{\lambda}^T.
  \label{eq:homogeneous}
\end{equation}
Then the optimization in Eq.~\eqref{eq:relaxationOptimizationOneWayCCQudit} can
be simplified to
\begin{equation}
  \begin{aligned}
    &\maxover[w_{ij},~\rho_{ij}] \quad &&\min\{1-w_{ij},1-w\}\\
    &\subto  &&w\ge 0,~~w_{ij}\ge 0,\FA i\ne j\\
    &        &&w_{ij}w_{ji}\ge(1-w)^2\lambda_i^2\lambda_j^2,\FA i\ne j\\
    &        &&\sum_{j\ne i}w_{ij}=(1-w)(1-\lambda_i^2),\FA i.
  \end{aligned}
  \label{eq:simplifiedOptimizationOneWayCCQudit}
\end{equation}
The second constraint in Eq.~\eqref{eq:simplifiedOptimizationOneWayCCQudit} implies
\begin{equation}
  \sum_{j\ne i}w_{ij}w_{ji}\ge(1-w)^2\lambda_i^2(1-\lambda_i^2),\FA i.
  \label{eq:sumPPT}
\end{equation}
Combining Eq.~\eqref{eq:sumPPT} with the last constraint
in Eq.~\eqref{eq:simplifiedOptimizationOneWayCCQudit}, we get
\begin{equation}
  \max_{j\ne i}w_{ji}
  \ge\frac{\sum_{j\ne i}w_{ij}w_{ji}}{\sum_{j\ne i}w_{ij}}
  \ge(1-w)\lambda_i^2,\FA i.
  \label{eq:simplifiedConstrants}
\end{equation}
Note that the denominator $\sum_{j\ne i}w_{ij}=1-\rho_{ii}$ is always non-zero
when $d\ge 2$. Hence, a further relaxation of the optimization for
$v(\Omega^\rightarrow)$ can be obtained via
\begin{equation}
  \begin{aligned}
    &\maxover[w_{ij},~\rho_{ij}] \quad &&\min\{1-w_{ij},1-w\}\\
    &\subto  &&w\ge 0,~~w_{ij}\ge 0,\FA i\ne j\\
    &        &&\max_{j\ne i}w_{ji}\ge(1-w)\lambda_i^2,\FA i,
  \end{aligned}
  \label{eq:furtherSimplifiedOptimizationOneWayCCQuditA}
\end{equation}
which can be directly solved with the solution given by
\begin{equation}
  \max_{\Omega^\rightarrow}v(\Omega^\rightarrow)\le\frac{1}{1+\lambda_1^2},
  \label{eq:simplifiedOptimalSolutionOneWayA}
\end{equation}
where $\lambda_1$ is the largest Schmidt coefficient of $\ket{\psi}$. This is
exactly what we want to prove, namely,
Eq.~\eqref{eq:simplifiedOptimalSolutionOneWay}.

Now, we consider the general case and show that the optimal solution is still
given by Eq.~\eqref{eq:simplifiedOptimalSolutionOneWayA}.
In general, $\rho$ can be written as
\begin{equation}
  \rho=wA+(1-w)\vect{\lambda}\vect{\lambda}^T,
  \label{eq:nonhomogeneous}
\end{equation}
where $A$ is a Hermitian matrix and
$w=\norm{\rho-\vect{\lambda}\vect{\lambda}^T}$. Then, the constraint
$\rho\vect{\lambda}=\vect{\lambda}$ implies
\begin{equation}
  A\le\I,\quad A\vect{\lambda}=\vect{\lambda}.
  \label{eq:constraintsA}
\end{equation}
As we have shown for the special case in the main text, as long as
Eq.~\eqref{eq:simplifiedConstrants} holds, all the rest arguments follow
immediately. From Eqs.~\eqref{eq:nonhomogeneous} and
\eqref{eq:constraintsA}, we can show
\begin{equation}
  \begin{aligned}
    &\sum_{j=1}^d\abs{\rho_{ij}}^2=(\rho^2)_{ii}=w^2(A^2)_{ii}
    +(1-w^2)\lambda_i^2,\\
    &\rho_{ii}=wA_{ii}+(1-w)\lambda_i^2,\FA i.
  \end{aligned}
  \label{eq:rhoij}
\end{equation}
Combining Eq.~\eqref{eq:rhoij} with the constraints in
Eq.~\eqref{eq:relaxationOptimizationOneWayCCQudit}, we get, for any fixed $i$,
\begin{equation}
  \begin{aligned}
    \max_{j\ne i}w_{ji}&\ge\frac{\sum_{j\ne i}w_{ij}w_{ji}}{\sum_{j\ne
    i}w_{ij}}\ge\frac{\sum_{j=1}^d\abs{\rho_{ij}}^2-(\rho_{ii})^2}{1-\rho_{ii}}\\
    &\ge(1-w)\lambda_i^2\times\frac{[(1+w)-(1-w)\lambda_i^2]-2wA_{ii}}
    {[1-(1-w)\lambda_i^2]-wA_{ii}}\\
    &\ge(1-w)\lambda_i^2,
  \end{aligned}
  \label{eq:simplifiedConstrantsGeneral}
\end{equation}
where we have used the relation
$(A_{ii})^2\le(A^2)_{ii}=\sum_{j=1}^d\abs{A_{ij}}^2$ for the third inequality,
and $A_{ii}\le 1$ for the last one. Thus we get back
Eq.~\eqref{eq:simplifiedConstrants} as well as the relaxation
\eqref{eq:furtherSimplifiedOptimizationOneWayCCQuditA},
then the upper bound in Eq.~\eqref{eq:simplifiedOptimalSolutionOneWayA} follows
straightforwardly.

At last, we show that the constructed $\Omega^\rightarrow$ in
Eq.~\eqref{eq:oneWayPVMQudit}, which we restate below
\begin{equation}
  \Omega^\rightarrow=wP_{ZZ}+\frac{1-w}{\abs{G}}\sum_{g\in
  G}gX_\psi^\rightarrow g^\dagger,
  \label{eq:oneWayPVMQuditA}
\end{equation}
can achieve the upper bound in Eq.~\eqref{eq:simplifiedOptimalSolutionOneWayA}
when $w$ is suitably chosen. From Eqs.~\eqref{eq:fourierBasis} and
\eqref{eq:invariantBasis}, we get
\begin{equation}
  \begin{aligned}
    w_{ij}&=\bra{ij}\Omega^\rightarrow\ket{ij}=w\bra{ij}P_{ZZ}\ket{ij}
    +(1-w)\bra{ij}X_\psi^\rightarrow\ket{ij}\\
    &=(1-w)\lambda_j^2,\qquad\FA i\ne j,\\
    \rho_{ij}&=\bra{ii}\Omega^\rightarrow\ket{jj}=w\bra{ii}P_{ZZ}\ket{jj}
    +(1-w)\bra{ii}X_\psi^\rightarrow\ket{jj}\\
    &=(1-w)\lambda_i\lambda_j,\qquad\FA i\ne j.\\
    \rho_{ii}&=\bra{ii}\Omega^\rightarrow\ket{ii}=w\bra{ii}P_{ZZ}\ket{ii}
    +(1-w)\bra{ii}X_\psi^\rightarrow\ket{ii}\\
    &=w+(1-w)\lambda_i^2,\qquad\FA i.
  \end{aligned}
  \label{eq:coefficientsFourier}
\end{equation}
Then it can be easily seen that $\rho$ takes the form in
Eq.~\eqref{eq:homogeneous}. When $w$ is chosen as
\begin{equation}
  w=\frac{\lambda_1^2}{1+\lambda_1^2},
  \label{eq:wOneWay}
\end{equation}
where $\lambda_1$ is the largest Schmidt coefficient of $\ket{\psi}$, we
can directly show
\begin{equation}
  v(\Omega^\rightarrow)=\min\{1-w_{ij},1-w\}=\frac{1}{1+\lambda_1^2},
  \label{eq:optimalBoundOneWay}
\end{equation}
which is the upper bound in Eq.~\eqref{eq:simplifiedOptimalSolutionOneWayA}.

\section{Appendix C: Optimal two-way strategy}

\begin{figure}
  \centering
  \includegraphics[width=.8\linewidth]{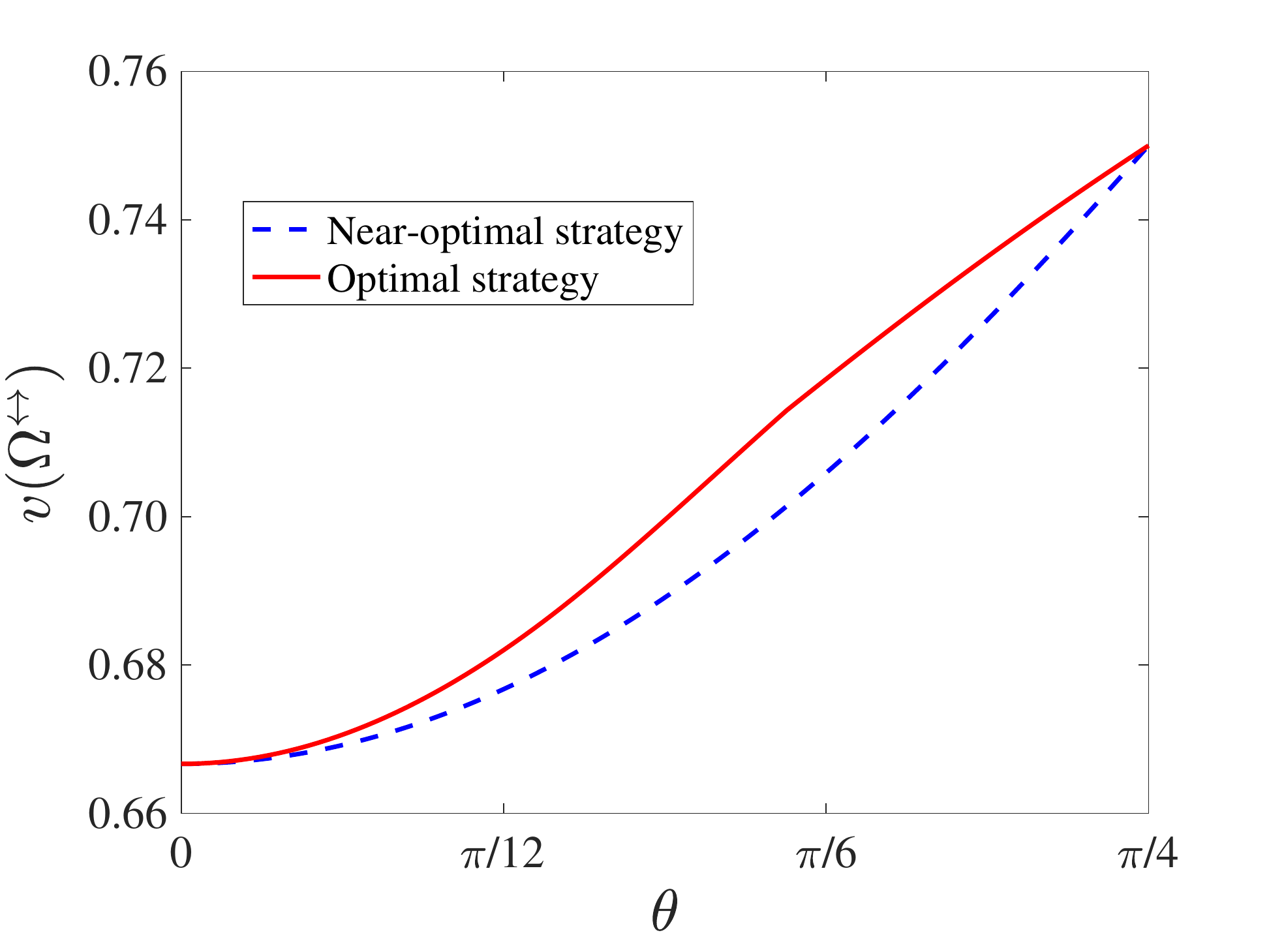}
  \caption{Comparison of the verification efficiency between the two-way optimal strategy
    from Eq.~\eqref{eq:relaxationOptimizationTwoWayCCQudit} and the two-way
    near-optimal strategy in Eq.~\eqref{eq:nearOptimalSolutionTwoWay} for
    two-qutrit pure states
    ${\ket{\psi}=\sqrt{\frac{2}{3}}\cos\theta\ket{00}+\sqrt{\frac{1}{3}}\ket{11}
    +\sqrt{\frac{2}{3}}\sin\theta\ket{22}}$ with $0\le\theta\le\pi/4$.  As we
    can see, the optimal efficiency is only slightly better than the
    near-optimal efficiency. And the resulting optimal strategies
    $\Omega^{\leftrightarrow}$ are verified to be separable with the method in
  Ref.~\cite{Shang.Guehne2018}.}
  \label{fig:twoWay}
\end{figure}

Similar to the one-way scenario, we can get a relaxation of the
optimization in Eq.~\eqref{eq:optimizationTwoWayGeneral} for the optimal
two-way strategy, which reads
\begin{equation}
  \begin{aligned}
    &\maxover[w_{ij},~\rho_{ij}]\quad&&\min\left\{1-\tfrac{1}{2}(w_{ij}+w_{ji}),
    1-\norm{\rho-\vect{\lambda}\vect{\lambda}^T}\right\},\\
    &\subto  &&0\le\rho\le\I,~~w_{ij}\ge 0,\FA i\ne j,\\
    &        &&w_{ij}w_{ji}\ge\abs{\rho_{ij}}^2,\FA i\ne j,\\
    &        &&\sum_{j\ne i}w_{ij}+\rho_{ii}=1,\FA i,\\
    &        &&\rho\vect{\lambda}=\vect{\lambda}.
  \end{aligned}
  \label{eq:relaxationOptimizationTwoWayCCQudit}
\end{equation}
This optimization, however, cannot be solved analytically.  Instead, we resort
to a numerical approach, then confirm the separability of the resulting
strategies with the method in Ref.~\cite{Shang.Guehne2018}.  See
Fig.~\ref{fig:twoWay} for a comparison of the verification efficiency between
the optimal strategy from Eq.~\eqref{eq:relaxationOptimizationTwoWayCCQudit}
and the near-optimal strategy in Eq.~\eqref{eq:nearOptimalSolutionTwoWay} for
two-qutrit states. As can be seen, the optimal efficiency is only slightly
better than the near-optimal efficiency, whereas the measurement settings of
the optimal strategy can be more complicated. Similar conclusions are also observed in
higher-dimensional cases. For instance, we have tested one million
randomly-drawn states for $d\le 10$. The results show that the optimal strategy
is at most $4\%$ better in efficiency than the near-optimal strategy.


%
\bibliography{QuantumInf}

\begin{thebibliography}{27}%
\makeatletter
\providecommand \@ifxundefined [1]{%
 \@ifx{#1\undefined}
}%
\providecommand \@ifnum [1]{%
 \ifnum #1\expandafter \@firstoftwo
 \else \expandafter \@secondoftwo
 \fi
}%
\providecommand \@ifx [1]{%
 \ifx #1\expandafter \@firstoftwo
 \else \expandafter \@secondoftwo
 \fi
}%
\providecommand \natexlab [1]{#1}%
\providecommand \enquote  [1]{``#1''}%
\providecommand \bibnamefont  [1]{#1}%
\providecommand \bibfnamefont [1]{#1}%
\providecommand \citenamefont [1]{#1}%
\providecommand \href@noop [0]{\@secondoftwo}%
\providecommand \href [0]{\begingroup \@sanitize@url \@href}%
\providecommand \@href[1]{\@@startlink{#1}\@@href}%
\providecommand \@@href[1]{\endgroup#1\@@endlink}%
\providecommand \@sanitize@url [0]{\catcode `\\12\catcode `\$12\catcode
  `\&12\catcode `\#12\catcode `\^12\catcode `\_12\catcode `\%12\relax}%
\providecommand \@@startlink[1]{}%
\providecommand \@@endlink[0]{}%
\providecommand \url  [0]{\begingroup\@sanitize@url \@url }%
\providecommand \@url [1]{\endgroup\@href {#1}{\urlprefix }}%
\providecommand \urlprefix  [0]{URL }%
\providecommand \Eprint [0]{\href }%
\providecommand \doibase [0]{http://dx.doi.org/}%
\providecommand \selectlanguage [0]{\@gobble}%
\providecommand \bibinfo  [0]{\@secondoftwo}%
\providecommand \bibfield  [0]{\@secondoftwo}%
\providecommand \translation [1]{[#1]}%
\providecommand \BibitemOpen [0]{}%
\providecommand \bibitemStop [0]{}%
\providecommand \bibitemNoStop [0]{.\EOS\space}%
\providecommand \EOS [0]{\spacefactor3000\relax}%
\providecommand \BibitemShut  [1]{\csname bibitem#1\endcsname}%
\let\auto@bib@innerbib\@empty
\bibitem [{\citenamefont {Paris}\ and\ \citenamefont
  {\v{R}eh\'a\v{c}ek}(2004)}]{Paris.Rehacek2004}%
  \BibitemOpen
  \bibinfo {editor} {\bibfnamefont {M.}~\bibnamefont {Paris}}\ and\ \bibinfo
  {editor} {\bibfnamefont {J.}~\bibnamefont {\v{R}eh\'a\v{c}ek}},\ eds.,\
  \href@noop {} {\emph {\bibinfo {title} {Quantum State Estimation}}},\
  \bibinfo {series} {Lecture Notes in Physics}, Vol.\ \bibinfo {volume} {649}\
  (\bibinfo  {publisher} {Springer},\ \bibinfo {address} {Heidelberg},\
  \bibinfo {year} {2004})\BibitemShut {NoStop}%
\bibitem [{\citenamefont {H{\"a}ffner}\ \emph {et~al.}(2005)\citenamefont
  {H{\"a}ffner}, \citenamefont {H{\"a}nsel}, \citenamefont {Roos},
  \citenamefont {Benhelm}, \citenamefont {{Chek-al-kar}}, \citenamefont
  {Chwalla}, \citenamefont {K{\"o}rber}, \citenamefont {Rapol}, \citenamefont
  {Riebe}, \citenamefont {Schmidt}, \citenamefont {Becher}, \citenamefont
  {G\"uhne}, \citenamefont {D\"ur},\ and\ \citenamefont
  {Blatt}}]{Haeffner.etal2005}%
  \BibitemOpen
  \bibfield  {author} {\bibinfo {author} {\bibfnamefont {H.}~\bibnamefont
  {H{\"a}ffner}}, \bibinfo {author} {\bibfnamefont {W.}~\bibnamefont
  {H{\"a}nsel}}, \bibinfo {author} {\bibfnamefont {C.~F.}\ \bibnamefont
  {Roos}}, \bibinfo {author} {\bibfnamefont {J.}~\bibnamefont {Benhelm}},
  \bibinfo {author} {\bibfnamefont {D.}~\bibnamefont {{Chek-al-kar}}}, \bibinfo
  {author} {\bibfnamefont {M.}~\bibnamefont {Chwalla}}, \bibinfo {author}
  {\bibfnamefont {T.}~\bibnamefont {K{\"o}rber}}, \bibinfo {author}
  {\bibfnamefont {U.~D.}\ \bibnamefont {Rapol}}, \bibinfo {author}
  {\bibfnamefont {M.}~\bibnamefont {Riebe}}, \bibinfo {author} {\bibfnamefont
  {P.~O.}\ \bibnamefont {Schmidt}}, \bibinfo {author} {\bibfnamefont
  {C.}~\bibnamefont {Becher}}, \bibinfo {author} {\bibfnamefont
  {O.}~\bibnamefont {G\"uhne}}, \bibinfo {author} {\bibfnamefont
  {W.}~\bibnamefont {D\"ur}}, \ and\ \bibinfo {author} {\bibfnamefont
  {R.}~\bibnamefont {Blatt}},\ }\href {\doibase 10.1038/nature04279} {\bibfield
   {journal} {\bibinfo  {journal} {Nature}\ }\textbf {\bibinfo {volume}
  {438}},\ \bibinfo {pages} {643} (\bibinfo {year} {2005})}\BibitemShut
  {NoStop}%
\bibitem [{\citenamefont {Shang}\ \emph {et~al.}(2017)\citenamefont {Shang},
  \citenamefont {Zhang},\ and\ \citenamefont {Ng}}]{Shang.etal2017}%
  \BibitemOpen
  \bibfield  {author} {\bibinfo {author} {\bibfnamefont {J.}~\bibnamefont
  {Shang}}, \bibinfo {author} {\bibfnamefont {Z.}~\bibnamefont {Zhang}}, \ and\
  \bibinfo {author} {\bibfnamefont {H.~K.}\ \bibnamefont {Ng}},\ }\href
  {\doibase 10.1103/PhysRevA.95.062336} {\bibfield  {journal} {\bibinfo
  {journal} {Phys. Rev. A}\ }\textbf {\bibinfo {volume} {95}},\ \bibinfo
  {pages} {062336} (\bibinfo {year} {2017})}\BibitemShut {NoStop}%
\bibitem [{\citenamefont {Schwemmer}\ \emph {et~al.}(2015)\citenamefont
  {Schwemmer}, \citenamefont {Knips}, \citenamefont {Richart}, \citenamefont
  {Weinfurter}, \citenamefont {Moroder}, \citenamefont {Kleinmann},\ and\
  \citenamefont {G\"uhne}}]{Schwemmer.etal2015}%
  \BibitemOpen
  \bibfield  {author} {\bibinfo {author} {\bibfnamefont {C.}~\bibnamefont
  {Schwemmer}}, \bibinfo {author} {\bibfnamefont {L.}~\bibnamefont {Knips}},
  \bibinfo {author} {\bibfnamefont {D.}~\bibnamefont {Richart}}, \bibinfo
  {author} {\bibfnamefont {H.}~\bibnamefont {Weinfurter}}, \bibinfo {author}
  {\bibfnamefont {T.}~\bibnamefont {Moroder}}, \bibinfo {author} {\bibfnamefont
  {M.}~\bibnamefont {Kleinmann}}, \ and\ \bibinfo {author} {\bibfnamefont
  {O.}~\bibnamefont {G\"uhne}},\ }\href {\doibase
  10.1103/PhysRevLett.114.080403} {\bibfield  {journal} {\bibinfo  {journal}
  {Phys. Rev. Lett.}\ }\textbf {\bibinfo {volume} {114}},\ \bibinfo {pages}
  {080403} (\bibinfo {year} {2015})}\BibitemShut {NoStop}%
\bibitem [{\citenamefont {T\'oth}\ and\ \citenamefont
  {G\"uhne}(2005)}]{Toth.Guehne2005c}%
  \BibitemOpen
  \bibfield  {author} {\bibinfo {author} {\bibfnamefont {G.}~\bibnamefont
  {T\'oth}}\ and\ \bibinfo {author} {\bibfnamefont {O.}~\bibnamefont
  {G\"uhne}},\ }\href {\doibase 10.1103/PhysRevLett.94.060501} {\bibfield
  {journal} {\bibinfo  {journal} {Phys. Rev. Lett.}\ }\textbf {\bibinfo
  {volume} {94}},\ \bibinfo {pages} {060501} (\bibinfo {year}
  {2005})}\BibitemShut {NoStop}%
\bibitem [{\citenamefont {Gross}\ \emph {et~al.}(2010)\citenamefont {Gross},
  \citenamefont {Liu}, \citenamefont {Flammia}, \citenamefont {Becker},\ and\
  \citenamefont {Eisert}}]{Gross.etal2010}%
  \BibitemOpen
  \bibfield  {author} {\bibinfo {author} {\bibfnamefont {D.}~\bibnamefont
  {Gross}}, \bibinfo {author} {\bibfnamefont {Y.-K.}\ \bibnamefont {Liu}},
  \bibinfo {author} {\bibfnamefont {S.~T.}\ \bibnamefont {Flammia}}, \bibinfo
  {author} {\bibfnamefont {S.}~\bibnamefont {Becker}}, \ and\ \bibinfo {author}
  {\bibfnamefont {J.}~\bibnamefont {Eisert}},\ }\href {\doibase
  10.1103/PhysRevLett.105.150401} {\bibfield  {journal} {\bibinfo  {journal}
  {Phys. Rev. Lett.}\ }\textbf {\bibinfo {volume} {105}},\ \bibinfo {pages}
  {150401} (\bibinfo {year} {2010})}\BibitemShut {NoStop}%
\bibitem [{\citenamefont {Flammia}\ and\ \citenamefont
  {Liu}(2011)}]{Flammia.Liu2011}%
  \BibitemOpen
  \bibfield  {author} {\bibinfo {author} {\bibfnamefont {S.~T.}\ \bibnamefont
  {Flammia}}\ and\ \bibinfo {author} {\bibfnamefont {Y.-K.}\ \bibnamefont
  {Liu}},\ }\href {\doibase 10.1103/PhysRevLett.106.230501} {\bibfield
  {journal} {\bibinfo  {journal} {Phys. Rev. Lett.}\ }\textbf {\bibinfo
  {volume} {106}},\ \bibinfo {pages} {230501} (\bibinfo {year}
  {2011})}\BibitemShut {NoStop}%
\bibitem [{\citenamefont {da~Silva}\ \emph {et~al.}(2011)\citenamefont
  {da~Silva}, \citenamefont {Landon-Cardinal},\ and\ \citenamefont
  {Poulin}}]{daSilva.etal2011}%
  \BibitemOpen
  \bibfield  {author} {\bibinfo {author} {\bibfnamefont {M.~P.}\ \bibnamefont
  {da~Silva}}, \bibinfo {author} {\bibfnamefont {O.}~\bibnamefont
  {Landon-Cardinal}}, \ and\ \bibinfo {author} {\bibfnamefont {D.}~\bibnamefont
  {Poulin}},\ }\href {\doibase 10.1103/PhysRevLett.107.210404} {\bibfield
  {journal} {\bibinfo  {journal} {Phys. Rev. Lett.}\ }\textbf {\bibinfo
  {volume} {107}},\ \bibinfo {pages} {210404} (\bibinfo {year}
  {2011})}\BibitemShut {NoStop}%
\bibitem [{\citenamefont {Pallister}\ \emph {et~al.}(2018)\citenamefont
  {Pallister}, \citenamefont {Linden},\ and\ \citenamefont
  {Montanaro}}]{Pallister.etal2018}%
  \BibitemOpen
  \bibfield  {author} {\bibinfo {author} {\bibfnamefont {S.}~\bibnamefont
  {Pallister}}, \bibinfo {author} {\bibfnamefont {N.}~\bibnamefont {Linden}}, \
  and\ \bibinfo {author} {\bibfnamefont {A.}~\bibnamefont {Montanaro}},\ }\href
  {\doibase 10.1103/PhysRevLett.120.170502} {\bibfield  {journal} {\bibinfo
  {journal} {Phys. Rev. Lett.}\ }\textbf {\bibinfo {volume} {120}},\ \bibinfo
  {pages} {170502} (\bibinfo {year} {2018})}\BibitemShut {NoStop}%
\bibitem [{\citenamefont {Dimi{\'c}}\ and\ \citenamefont
  {Daki{\'c}}(2018)}]{Dimic.Dakic2018}%
  \BibitemOpen
  \bibfield  {author} {\bibinfo {author} {\bibfnamefont {A.}~\bibnamefont
  {Dimi{\'c}}}\ and\ \bibinfo {author} {\bibfnamefont {B.}~\bibnamefont
  {Daki{\'c}}},\ }\href {\doibase 10.1038/s41534-017-0055-x} {\bibfield
  {journal} {\bibinfo  {journal} {npj Quantum Inf.}\ }\textbf {\bibinfo
  {volume} {4}},\ \bibinfo {pages} {11} (\bibinfo {year} {2018})}\BibitemShut
  {NoStop}%
\bibitem [{\citenamefont {Morimae}\ \emph {et~al.}(2017)\citenamefont
  {Morimae}, \citenamefont {Takeuchi},\ and\ \citenamefont
  {Hayashi}}]{Morimae.etal2017}%
  \BibitemOpen
  \bibfield  {author} {\bibinfo {author} {\bibfnamefont {T.}~\bibnamefont
  {Morimae}}, \bibinfo {author} {\bibfnamefont {Y.}~\bibnamefont {Takeuchi}}, \
  and\ \bibinfo {author} {\bibfnamefont {M.}~\bibnamefont {Hayashi}},\ }\href
  {\doibase 10.1103/PhysRevA.96.062321} {\bibfield  {journal} {\bibinfo
  {journal} {Phys. Rev. A}\ }\textbf {\bibinfo {volume} {96}},\ \bibinfo
  {pages} {062321} (\bibinfo {year} {2017})}\BibitemShut {NoStop}%
\bibitem [{\citenamefont {Takeuchi}\ and\ \citenamefont
  {Morimae}(2018)}]{Takeuchi.Morimae2018}%
  \BibitemOpen
  \bibfield  {author} {\bibinfo {author} {\bibfnamefont {Y.}~\bibnamefont
  {Takeuchi}}\ and\ \bibinfo {author} {\bibfnamefont {T.}~\bibnamefont
  {Morimae}},\ }\href {\doibase 10.1103/PhysRevX.8.021060} {\bibfield
  {journal} {\bibinfo  {journal} {Phys. Rev. X}\ }\textbf {\bibinfo {volume}
  {8}},\ \bibinfo {pages} {021060} (\bibinfo {year} {2018})}\BibitemShut
  {NoStop}%
\bibitem [{\citenamefont {Zhu}\ and\ \citenamefont
  {Hayashi}(2019{\natexlab{a}})}]{Zhu.Hayashi2018}%
  \BibitemOpen
  \bibfield  {author} {\bibinfo {author} {\bibfnamefont {H.}~\bibnamefont
  {Zhu}}\ and\ \bibinfo {author} {\bibfnamefont {M.}~\bibnamefont {Hayashi}},\
  }\href {\doibase 10.1103/PhysRevApplied.12.054047} {\bibfield  {journal}
  {\bibinfo  {journal} {Phys. Rev. Applied}\ }\textbf {\bibinfo {volume}
  {12}},\ \bibinfo {pages} {054047} (\bibinfo {year}
  {2019}{\natexlab{a}})}\BibitemShut {NoStop}%
\bibitem [{\citenamefont {Chernoff}(1952)}]{Chernoff1952}%
  \BibitemOpen
  \bibfield  {author} {\bibinfo {author} {\bibfnamefont {H.}~\bibnamefont
  {Chernoff}},\ }\href {\doibase 10.1214/aoms/1177729330} {\bibfield  {journal}
  {\bibinfo  {journal} {Ann. Math. Statist.}\ }\textbf {\bibinfo {volume}
  {23}},\ \bibinfo {pages} {493} (\bibinfo {year} {1952})}\BibitemShut
  {NoStop}%
\bibitem [{\citenamefont {Hoeffding}(1963)}]{Hoeffding1963}%
  \BibitemOpen
  \bibfield  {author} {\bibinfo {author} {\bibfnamefont {W.}~\bibnamefont
  {Hoeffding}},\ }\href {\doibase 10.1080/01621459.1963.10500830} {\bibfield
  {journal} {\bibinfo  {journal} {J. Am. Stat. Assoc.}\ }\textbf {\bibinfo
  {volume} {58}},\ \bibinfo {pages} {13} (\bibinfo {year} {1963})}\BibitemShut
  {NoStop}%
\bibitem [{\citenamefont {Bennett}\ \emph {et~al.}(1996)\citenamefont
  {Bennett}, \citenamefont {Brassard}, \citenamefont {Popescu}, \citenamefont
  {Schumacher}, \citenamefont {Smolin},\ and\ \citenamefont
  {Wootters}}]{Bennett.etal1996}%
  \BibitemOpen
  \bibfield  {author} {\bibinfo {author} {\bibfnamefont {C.~H.}\ \bibnamefont
  {Bennett}}, \bibinfo {author} {\bibfnamefont {G.}~\bibnamefont {Brassard}},
  \bibinfo {author} {\bibfnamefont {S.}~\bibnamefont {Popescu}}, \bibinfo
  {author} {\bibfnamefont {B.}~\bibnamefont {Schumacher}}, \bibinfo {author}
  {\bibfnamefont {J.~A.}\ \bibnamefont {Smolin}}, \ and\ \bibinfo {author}
  {\bibfnamefont {W.~K.}\ \bibnamefont {Wootters}},\ }\href {\doibase
  10.1103/PhysRevLett.76.722} {\bibfield  {journal} {\bibinfo  {journal} {Phys.
  Rev. Lett.}\ }\textbf {\bibinfo {volume} {76}},\ \bibinfo {pages} {722}
  (\bibinfo {year} {1996})}\BibitemShut {NoStop}%
\bibitem [{\citenamefont {Watrous}(2018)}]{Watrous2018}%
  \BibitemOpen
  \bibfield  {author} {\bibinfo {author} {\bibfnamefont {J.}~\bibnamefont
  {Watrous}},\ }\href@noop {} {\emph {\bibinfo {title} {The Theory of Quantum
  Information}}}\ (\bibinfo  {publisher} {Cambridge University Press,
  Cambridge, UK},\ \bibinfo {year} {2018})\BibitemShut {NoStop}%
\bibitem [{\citenamefont {Peres}(2002)}]{Peres2002}%
  \BibitemOpen
  \bibfield  {author} {\bibinfo {author} {\bibfnamefont {A.}~\bibnamefont
  {Peres}},\ }\href@noop {} {\emph {\bibinfo {title} {Quantum Theory: Concepts
  and Methods}}}\ (\bibinfo  {publisher} {Kluwer Academic Publishers},\
  \bibinfo {year} {2002})\BibitemShut {NoStop}%
\bibitem [{\citenamefont {Peres}(1996)}]{Peres1996}%
  \BibitemOpen
  \bibfield  {author} {\bibinfo {author} {\bibfnamefont {A.}~\bibnamefont
  {Peres}},\ }\href {\doibase 10.1103/PhysRevLett.77.1413} {\bibfield
  {journal} {\bibinfo  {journal} {Phys. Rev. Lett.}\ }\textbf {\bibinfo
  {volume} {77}},\ \bibinfo {pages} {1413} (\bibinfo {year}
  {1996})}\BibitemShut {NoStop}%
\bibitem [{\citenamefont {Horodecki}\ \emph {et~al.}(1996)\citenamefont
  {Horodecki}, \citenamefont {Horodecki},\ and\ \citenamefont
  {Horodecki}}]{Horodecki.etal1996}%
  \BibitemOpen
  \bibfield  {author} {\bibinfo {author} {\bibfnamefont {M.}~\bibnamefont
  {Horodecki}}, \bibinfo {author} {\bibfnamefont {P.}~\bibnamefont
  {Horodecki}}, \ and\ \bibinfo {author} {\bibfnamefont {R.}~\bibnamefont
  {Horodecki}},\ }\href {\doibase 10.1016/S0375-9601(96)00706-2} {\bibfield
  {journal} {\bibinfo  {journal} {Phys. Lett. A}\ }\textbf {\bibinfo {volume}
  {223}},\ \bibinfo {pages} {1} (\bibinfo {year} {1996})}\BibitemShut {NoStop}%
\bibitem [{\citenamefont {Zhu}\ and\ \citenamefont
  {Hayashi}(2019{\natexlab{b}})}]{Zhu.Hayashi2019}%
  \BibitemOpen
  \bibfield  {author} {\bibinfo {author} {\bibfnamefont {H.}~\bibnamefont
  {Zhu}}\ and\ \bibinfo {author} {\bibfnamefont {M.}~\bibnamefont {Hayashi}},\
  }\href {\doibase 10.1103/PhysRevA.99.052346} {\bibfield  {journal} {\bibinfo
  {journal} {Phys. Rev. A}\ }\textbf {\bibinfo {volume} {99}},\ \bibinfo
  {pages} {052346} (\bibinfo {year} {2019}{\natexlab{b}})}\BibitemShut
  {NoStop}%
\bibitem [{\citenamefont {Shang}\ and\ \citenamefont
  {G\"uhne}(2018)}]{Shang.Guehne2018}%
  \BibitemOpen
  \bibfield  {author} {\bibinfo {author} {\bibfnamefont {J.}~\bibnamefont
  {Shang}}\ and\ \bibinfo {author} {\bibfnamefont {O.}~\bibnamefont
  {G\"uhne}},\ }\href {\doibase 10.1103/PhysRevLett.120.050506} {\bibfield
  {journal} {\bibinfo  {journal} {Phys. Rev. Lett.}\ }\textbf {\bibinfo
  {volume} {120}},\ \bibinfo {pages} {050506} (\bibinfo {year}
  {2018})}\BibitemShut {NoStop}%
\bibitem [{\citenamefont {Liu}\ \emph {et~al.}(2019)\citenamefont {Liu},
  \citenamefont {Yu}, \citenamefont {Shang}, \citenamefont {Zhu},\ and\
  \citenamefont {Zhang}}]{Liu.etal2019}%
  \BibitemOpen
  \bibfield  {author} {\bibinfo {author} {\bibfnamefont {Y.-C.}\ \bibnamefont
  {Liu}}, \bibinfo {author} {\bibfnamefont {X.-D.}\ \bibnamefont {Yu}},
  \bibinfo {author} {\bibfnamefont {J.}~\bibnamefont {Shang}}, \bibinfo
  {author} {\bibfnamefont {H.}~\bibnamefont {Zhu}}, \ and\ \bibinfo {author}
  {\bibfnamefont {X.}~\bibnamefont {Zhang}},\ }\href {\doibase
  10.1103/PhysRevApplied.12.044020} {\bibfield  {journal} {\bibinfo  {journal}
  {Phys. Rev. Applied}\ }\textbf {\bibinfo {volume} {12}},\ \bibinfo {pages}
  {044020} (\bibinfo {year} {2019})}\BibitemShut {NoStop}%
\bibitem [{\citenamefont {Wang}\ and\ \citenamefont
  {Hayashi}(2019)}]{Wang.Hayashi2019}%
  \BibitemOpen
  \bibfield  {author} {\bibinfo {author} {\bibfnamefont {K.}~\bibnamefont
  {Wang}}\ and\ \bibinfo {author} {\bibfnamefont {M.}~\bibnamefont {Hayashi}},\
  }\href {\doibase 10.1103/PhysRevA.100.032315} {\bibfield  {journal} {\bibinfo
   {journal} {Phys. Rev. A}\ }\textbf {\bibinfo {volume} {100}},\ \bibinfo
  {pages} {032315} (\bibinfo {year} {2019})}\BibitemShut {NoStop}%
\bibitem [{\citenamefont {Barnett}\ and\ \citenamefont
  {Croke}(2009)}]{Barnett.Croke2009}%
  \BibitemOpen
  \bibfield  {author} {\bibinfo {author} {\bibfnamefont {S.~M.}\ \bibnamefont
  {Barnett}}\ and\ \bibinfo {author} {\bibfnamefont {S.}~\bibnamefont
  {Croke}},\ }\href {\doibase 10.1364/AOP.1.000238} {\bibfield  {journal}
  {\bibinfo  {journal} {Adv. Opt. Photonics}\ }\textbf {\bibinfo {volume}
  {1}},\ \bibinfo {pages} {238} (\bibinfo {year} {2009})}\BibitemShut {NoStop}%
\bibitem [{\citenamefont {Bae}\ and\ \citenamefont
  {Kwek}(2015)}]{Bae.Kwek2015}%
  \BibitemOpen
  \bibfield  {author} {\bibinfo {author} {\bibfnamefont {J.}~\bibnamefont
  {Bae}}\ and\ \bibinfo {author} {\bibfnamefont {L.-C.}\ \bibnamefont {Kwek}},\
  }\href {\doibase 10.1088/1751-8113/48/8/083001} {\bibfield  {journal}
  {\bibinfo  {journal} {J. Phys. A: Math. Theor.}\ }\textbf {\bibinfo {volume}
  {48}},\ \bibinfo {pages} {083001} (\bibinfo {year} {2015})}\BibitemShut
  {NoStop}%
\bibitem [{\citenamefont {Li}\ \emph {et~al.}(2019)\citenamefont {Li},
  \citenamefont {Han},\ and\ \citenamefont {Zhu}}]{Li.etal2019}%
  \BibitemOpen
  \bibfield  {author} {\bibinfo {author} {\bibfnamefont {Z.}~\bibnamefont
  {Li}}, \bibinfo {author} {\bibfnamefont {Y.-G.}\ \bibnamefont {Han}}, \ and\
  \bibinfo {author} {\bibfnamefont {H.}~\bibnamefont {Zhu}},\ }\href {\doibase
  10.1103/PhysRevA.100.032316} {\bibfield  {journal} {\bibinfo  {journal}
  {Phys. Rev. A}\ }\textbf {\bibinfo {volume} {100}},\ \bibinfo {pages}
  {032316} (\bibinfo {year} {2019})}\BibitemShut {NoStop}%
\end{thebibliography}%

\end{document}